\documentclass[aps,pra,reprint,superscriptaddress,nofootinbib,longbibliography]{revtex4-2}

\usepackage{amsmath,amssymb}
\usepackage{graphicx}
\usepackage{bm}
\usepackage[colorlinks=true,linkcolor=blue,citecolor=blue,urlcolor=blue]{hyperref}

\newcommand{\ket}[1]{|#1\rangle}
\newcommand{\bra}[1]{\langle#1|}
\newcommand{\braket}[2]{\langle#1|#2\rangle}
\newcommand{\dmu}{d\mu}
\newcommand{\Pc}{P_{-1/2+i\lambda}}
\newcommand{\dOm}{d\Omega_\lambda}

\begin{document}

\title{Hudson's theorem fails for the
\texorpdfstring{$SU(1,1)$}{SU(1,1)} discrete series}

\author{Chon-Fai Kam}
\email{dubussygauss@gmail.com}
\affiliation{Dipartimento di Fisica e Chimica ``Emilio Segr\`e'',
Universit\`a degli Studi di Palermo, via Archirafi 36, I-90123 Palermo, Italy}
\affiliation{Universit\'e Paris Cit\'e and Universit\'e de La R\'eunion,
INSERM, BIGR, DSIMB, F-75015 Paris, France}

\date{\today}

\begin{abstract}
Hudson's theorem states that a pure state of a bosonic mode has a
non-negative Wigner function if and only if it is Gaussian.  It
underwrites the reading of Wigner negativity as a faithful signature of
pure-state non-classicality.  We show the statement has no analogue on
curved phase space.  For the positive discrete series of $SU(1,1)$,
realised on the upper sheet of a two-sheeted hyperboloid, the
Wigner-positive pure states form a strictly larger set than the Perelomov
coherent orbit.  Superpositions of the
lowest weight state with the first excited state stay positive up to a
mixing angle of $24.93^\circ$ at Bargmann index $k=1$, and the admissible
set has positive volume, with a maximal width that is not attained in the
two-state direction.  We show analytically that the quadratic form controlling
positivity degenerates in the far field onto a single mixing angle.  That
degeneracy bounds the window by $\arctan(1/\sqrt{2k})$ and leaves the
threshold itself fixed at intermediate hyperbolic distance.
\end{abstract}

\maketitle

\section{Introduction}
\label{sec:intro}

The Wigner map organises quantum mechanics as a statistical theory on
phase space~\cite{Groenewold1946,Moyal1949,Hillery1984}, and it sits in a
one-parameter family of quasiprobabilities labelled by an ordering
parameter~\cite{CahillGlauber1969}.  Within that family the Wigner case
is singled out by self-duality, which lets one expectation value be
computed from a single function rather than a dual pair.  The price is
that the function takes negative values on most states.  Hudson proves
that for a single bosonic mode the exceptions are exactly the Gaussian
pure states~\cite{Hudson1974}, and Soto and Claverie extend the statement
to $n$ modes~\cite{SotoClaverie1983}.

Hudson's theorem is best read as a statement about a distinguished class
rather than about Gaussians as such.  It says that the states with a
non-negative Wigner function coincide with the class the geometry
already singles out, and it therefore takes a different form in every
setting where that class is different.  Gross proves the
finite-dimensional counterpart, where stabiliser states play the role of
the Gaussians~\cite{Gross2006}, and Mandilara, Karpov and Cerf push the
continuous-variable version towards mixed states~\cite{Mandilara2009}.
What the theorem asserts in a given setting is that two definitions of
classicality agree, one supplied by the symmetry and one supplied by the
phase-space function.

A large part of the operational use of Wigner negativity rests on that
agreement.  Kenfack and \.Zyczkowski take the negative volume as an
indicator of non-classicality~\cite{Kenfack2004}.  Mari and Eisert bound
classical simulability by Wigner positivity~\cite{MariEisert2012}.
Chabaud and co-workers build the stellar hierarchy on the departure from
Gaussianity~\cite{Chabaud2020}, which Walschaers reviews together with
the rest of the non-Gaussian toolkit~\cite{Walschaers2021}.  Booth,
Chabaud and Emeriau prove that contextuality and Wigner negativity
coincide for continuous-variable measurement models
\cite{BoothChabaudEmeriau2022}.  Each of these results reads a property
of the function as a property of the state, and each is proved on the
flat phase space of the Heisenberg-Weyl group.

Phase space need not be flat.  The Stratonovich-Weyl construction
produces a covariant quasiprobability for any dynamical
symmetry~\cite{Stratonovich1956,Brif1999}, with the coherent states of
that symmetry labelling the manifold~\cite{Perelomov1986,Zhang1990,Kam2023},
and the group rather than the physicist fixes the geometry that carries
it.
The construction has been carried out for
spin~\cite{Agarwal1981,Varilly1989,Klimov2017}, for systems of arbitrary
dimension~\cite{Tilma2016}, and for hyperboloids~\cite{Alonso2002}.  In each case the symmetry supplies a natural candidate for the class of
states a phase-space treatment would call classical, namely the orbit of
coherent states.  The phase-space function supplies a second candidate,
namely the states whose Wigner function stays non-negative.  We call
either candidate a free class, borrowing the term from resource
theories, where it names the states a theory hands out for nothing.
Whether the two candidates still agree is a question about the geometry.

The question has become concrete for $SU(1,1)$.  The algebra governs the
non-degenerate parametric amplifier and the two-mode squeezed
states~\cite{Wodkiewicz1985,Gerry1985}.  It also governs the
interferometer of Yurke, McCall and Klauder~\cite{Yurke1986}, which has
been built with parametric amplifiers~\cite{Jing2011,Hudelist2014} and is
now standard in nonlinear interferometry~\cite{Chekhova2016}.  The matching phase-space
machinery arrived recently.  Seyfarth \emph{et al.} construct the
$SU(1,1)$ Wigner function~\cite{Seyfarth2020}, Klimov \emph{et al.} embed
it in a covariant $s$-parametrised family~\cite{Klimov2020} and work out
the differential form of the star product~\cite{Baltazar2025}, and Klimov
\emph{et al.} then prove that every Perelomov coherent state has a
strictly positive Wigner function~\cite{Klimov2026}.  An experiment that
reconstructs an $SU(1,1)$ Wigner function and finds it non-negative would
today be read as having prepared a coherent state.  We show that reading
is not available.

The hyperboloid is also the only constant-curvature phase space where the
question can be posed at all.  On the sphere every spin coherent state
has an $SU(2)$ Wigner function that takes negative values, and the same
is conjectured for every pure spin state~\cite{Davis2021,Davis2023}, so
the candidate free class is void or nearly so and there is nothing for
the two definitions to disagree about.  On the two-sheeted hyperboloid
the free class is non-empty by the result of Klimov \emph{et
al.}~\cite{Klimov2026}, and it is selected by the group rather than
assumed.  Hudson's dichotomy therefore has something to separate, and we
ask whether it does.

We answer in the negative.  To our knowledge the Wigner-positive pure
states on the hyperboloid have not been characterised before, and we
exhibit a whole family of them that lie off the coherent orbit.
Section~\ref{sec:setup} fixes notation.  Section~\ref{sec:blocks} reduces
the Wigner symbols in the number basis to a two-dimensional function
space, which turns the positivity question into linear algebra.
Section~\ref{sec:beyond} settles that question for the simplest
superposition, where the threshold turns out to be finite, and shows the
admissible states fill a region rather than a curve.  Section~\ref{sec:kdep} traces the
threshold to the far field, where the quadratic form degenerates onto a
single direction, and contrasts the outcome with the flat case.
Section~\ref{sec:disc} collects the consequences and the limitations.

\section{Phase space for \texorpdfstring{$SU(1,1)$}{SU(1,1)}}
\label{sec:setup}


The algebra $su(1,1)$ is spanned by $\{\hat K_0,\hat K_1,\hat K_2\}$ with
$[\hat K_1,\hat K_2]=-i\hat K_0$ and cyclic partners, and we write
$\hat K_\pm=\pm i(\hat K_1\pm i\hat K_2)$, so that
\begin{equation}
  [\hat K_0,\hat K_\pm]=\pm \hat K_\pm ,
  \qquad
  [\hat K_-,\hat K_+]=2\hat K_0 .
\end{equation}
We work in the positive discrete series, labelled by the Bargmann index
$k=\tfrac12,1,\tfrac32,\dots$, and spanned by $\ket{k,k+m}$ with
\begin{equation}
  \hat K_0\ket{k,k+m}=(k+m)\ket{k,k+m},
  \qquad
  \hat K_-\ket{k,k}=0 .
\end{equation}
The coherent states resolve the identity for $k>1/2$~\cite{Klimov2020},
and all results below are quoted for $k\ge1$, which also keeps us clear
of the two irreps $k=\tfrac14$ and $k=\tfrac34$ carried by the
single-mode oscillator.

The realisation that motivates the choice is the non-degenerate
parametric amplifier of Sec.~\ref{sec:intro}~\cite[Eq.~(4.1)]{Klimov2020},
where $\hat K_+=\hat a^\dagger\hat b^\dagger$,
$\hat K_-=\hat a\hat b$ and
$\hat K_0=\tfrac12(\hat a^\dagger\hat a+\hat b^\dagger\hat b+1)$.  The
subspace with a fixed excitation difference $\Delta n$ between the two
modes carries the irrep with $k=\tfrac12(1+|\Delta n|)$, so the Bargmann
index is set by a quantity an experiment prepares.

The Perelomov coherent states are the orbit of the lowest weight
vector~\cite{Perelomov1986,Kam2023},
\begin{equation}
  \ket{\zeta}=(1-|\zeta|^2)^k\sum_{m\ge0}
  \sqrt{\frac{\Gamma(m+2k)}{m!\,\Gamma(2k)}}\;\zeta^m\ket{k,k+m},
  \label{eq:coh}
\end{equation}
with $\zeta=\tanh(\tau/2)\,e^{-i\phi}$ in the Poincar\'e disc.  Inverse
stereographic projection lifts the disc to the upper sheet of the
two-sheeted hyperboloid, which carries the invariant measure
\begin{equation}
  \dmu(\zeta)=\frac{d^2\zeta}{(1-|\zeta|^2)^2}
  =\tfrac14\sinh\tau\,d\tau\,d\phi .
  \label{eq:measure}
\end{equation}

We use the covariant kernels $\hat w^{(s)}(\zeta)$ of
Refs.~\cite{Seyfarth2020,Klimov2020}, normalised by
$\mathrm{Tr}\,\hat w^{(s)}(\zeta)=1$, with symbols
$W^{(s)}_A(\zeta)=\mathrm{Tr}[A\,\hat w^{(s)}(\zeta)]$.  The two ends of
the family are the familiar ones.  At $s=-1$ the symbol is the Husimi
function $Q_A(\zeta)=\bra{\zeta}A\ket{\zeta}$, and at $s=+1$ it is the
$P$ symbol defined by
$A=(2k-1)\pi^{-1}\!\int\dmu\,P_A\ket{\zeta}\bra{\zeta}$.  The value $s=0$
is self-dual,
\begin{equation}
  \mathrm{Tr}[AB]=\frac{2k-1}{\pi}\int \dmu(\zeta)\,
  W_A(\zeta)\,W_B(\zeta),
  \label{eq:selfdual}
\end{equation}
and we write $W_A \equiv W^{(0)}_A$ throughout.  A state of unit trace
satisfies $(2k-1)\pi^{-1}\!\int\dmu\,W_\rho=1$.

Two abbreviations recur,
\begin{equation}
  u=\cosh\tau,
  \qquad
  \nu=-\tfrac14-\lambda^2 ,
  \label{eq:abbrev}
\end{equation}
together with the function
\begin{equation}
  \Phi_k(\lambda)=\frac{2k-1}{\Gamma(2k)^2}\,
  \Bigl|\Gamma\!\left(2k-\tfrac12+i\lambda\right)\Bigr|^2 ,
  \label{eq:Phi}
\end{equation}
which obeys
\begin{equation}
  \frac{2}{2k-1}\int_0^\infty\! d\lambda\,\lambda\tanh(\pi\lambda)\,
  \Phi_k(\lambda)=1 .
  \label{eq:Phinorm}
\end{equation}
Every integral below runs over the same measure, so we abbreviate
\begin{equation}
  \dOm \equiv d\lambda\,\lambda\tanh(\pi\lambda)\,\Phi_k^{1/2}(\lambda).
  \label{eq:dOm}
\end{equation}
The kernel carries the Wigner symbol from the $P$ symbol by convolution
with a strictly positive profile, and the reverse convolution from the
$Q$ symbol is singular, so the $P$ route is the only practical
one~\cite{Klimov2020}.  We follow it throughout.

The conical function $\Pc$~\cite{Erdelyi1955,DLMF} plays the role that
plane waves play on flat phase space.  It is the zonal spherical function of the
hyperboloid~\cite[App.~A]{Klimov2020}, it depends on two points only
through their pseudo-scalar product, which we write as $u=\cosh\xi$ and
which reduces to $u=\cosh\tau$ when one of the points is the origin, and it diagonalises
the Laplace operator $\mathcal{L}^2$ of Eq.~(2.25) of
Ref.~\cite{Klimov2020},
\begin{equation}
  \mathcal{L}^2\Pc=\nu\,\Pc ,
  \qquad
  \nu=-\tfrac14-\lambda^2 .
  \label{eq:eigen}
\end{equation}
The abbreviation $\nu$ of Eq.~\eqref{eq:abbrev} is therefore the
Laplacian eigenvalue, and the parameter $\lambda$ labels the continuous
spectrum.  Equivalently $\Pc$ solves
\begin{equation}
  (u^2-1)P''+2uP'-\nu P=0 ,
  \label{eq:legendre}
\end{equation}
which we use repeatedly to eliminate higher derivatives.

The weight $\Phi_k$ enters through the same operator.  The squared
overlap of two coherent states is obtained from a delta function by
applying a group-invariant operator, which transitivity forces to be a
function of the Casimir~\cite[App.~A]{Klimov2020}, and
$\Phi_k(\mathcal{L}^2)$ is that function.  It carries the
$P$ kernel to the $Q$ kernel, so its square root carries the $P$ kernel
to the self-dual one and Eq.~\eqref{eq:Phi} is what appears under every
integral below.

The $P$ symbol of the lowest weight state is a delta function on the
hyperboloid~\cite[Eq.~(3.1)]{Klimov2020}, so its Wigner function follows
at once,
\begin{equation}
  W_{00}(\tau)=\frac{2}{2k-1}\int_0^\infty \dOm \; \Pc(\cosh\tau).
  \label{eq:W00}
\end{equation}
Covariance carries the same profile, evaluated at the hyperbolic distance
between two points, to every coherent state.  Klimov \emph{et al.} prove
it is strictly positive~\cite{Klimov2026}, and we confirm this
independently out to hyperbolic distance $20$ in
Appendix~\ref{app:valid}.

\section{Wigner symbols in the number basis}
\label{sec:blocks}

Testing Hudson's dichotomy requires the Wigner function of states outside
the coherent orbit, and the orbit is the one family where the answer is
already available.  Its $P$ symbol is a delta function on the
hyperboloid, so Eq.~\eqref{eq:W00} follows by inspection.  Every other
state needs the general machinery.

The number states $\ket{k,k+m}$ are the natural place to start.  They
diagonalise $\hat K_0$, every state in the irrep expands in them, and the
lowest of them is the coherent state we already control.  A superposition
of them needs more than the populations, since the Wigner function of
$\rho=\sum c_mc_n^*\ket{k,k+m}\bra{k,k+n}$ picks up a symbol from every
term.  We write $W_{mn}$ for the Wigner symbol of
$\ket{k,k+m}\bra{k,k+n}$ and call it a block, so that
$W_\rho=\sum c_mc_n^*W_{mn}$.

Klimov \emph{et al.} give the first excited state
explicitly~\cite[Eq.~(3.12)]{Klimov2020} and the general block as a
nested derivative of a generating function, reproduced as
Eq.~\eqref{eq:B6} below.  Evaluating that expression term by term is
awkward, because the differential operators involved carry a factor
$\coth(\tau'/2)$ that diverges at the origin, where the delta function
sets the argument.  Individual terms are then singular and only their
combination is finite.

The difficulty disappears once one notices that the answer always lands
in a two-dimensional function space.  The mechanism is that the conical
function obeys a second-order equation, so any polynomial in $d/du$
applied to it collapses modulo Eq.~\eqref{eq:legendre} to a first-order
expression.  Covariance under $\hat K_0$ fixes the angular dependence,
$W_{mn}\propto e^{-i(n-m)\phi}$, and once that factor and
$\tanh^{|m-n|}(\tau/2)$ are stripped, what remains is a linear
combination of $\Pc(u)$ and $\Pc'(u)$ with coefficients rational in $u$
and polynomial in $\nu$ and $k$.  Table~\ref{tab:blocks} lists the six
blocks used below, together with two more that the same reduction
produces, and Appendix~\ref{app:lemma} gives the derivation.

\begin{table}[tb]
\caption{\label{tab:blocks}Coefficients of the block reduction, with
$\nu=-\frac14-\lambda^2$, $u=\cosh\tau$ and $P\equiv\Pc(u)$.  The angular
factor $\tanh^{|m-n|}(\tau/2)\,e^{-i(n-m)\phi}$ has been stripped.}
\begin{ruledtabular}
\begin{tabular}{ll}
block & entry \\
\hline
$(0,0)$ & $P$ \\[3pt]
$(1,1)$ & $(2k+\nu)\,P$ \\[3pt]
$(2,2)$ & $\bigl(8k^2+8k\nu+4k+\nu^2+2\nu\bigr)P$ \\[3pt]
$(3,3)$ & $\bigl(48k^3+72k^2\nu+72k^2+18k\nu^2$ \\
        & $\quad+72k\nu+24k+\nu^3+10\nu^2+12\nu\bigr)P$ \\[3pt]
$(0,1)$ & $-(u+1)\,P'$ \\[3pt]
$(1,2)$ & $-(u+1)(4k+\nu)\,P'$ \\[3pt]
$(2,3)$ & $-(u+1)\bigl(24k^2+12k\nu+12k$ \\
        & $\qquad\quad+\nu^2+4\nu\bigr)P'$ \\[3pt]
$(0,2)$ & $\dfrac{u+1}{u-1}\bigl(\nu P-2u\,P'\bigr)$ \\[3pt]
\end{tabular}
\end{ruledtabular}
\end{table}

The pattern in Table~\ref{tab:blocks} is a selection rule.  Populations
see only $\Pc$, the nearest coherences see only $\Pc'$, and the two mix
only from the second coherence onwards.  The first case says that
$\hat K_0$-diagonal operators map to functions of the Laplacian on the
hyperboloid, which acts on $\Pc$ by multiplication by $\nu$.  Setting
$m=n=1$ and restoring the normalisation returns the coefficient
$2k-\frac14-\lambda^2$, which reproduces Eq.~(3.12) of
Ref.~\cite{Klimov2020} and is the consistency check we rely on
throughout.

Three consequences carry the rest of the paper.  Every state built from
finitely many number states has a Wigner function that is a finite linear
combination of two radial integrals, so the positivity question becomes
linear algebra with $u$-dependent coefficients.  All blocks share one
integral representation, which is what lets Sec.~\ref{sec:kdep} extract a
common exponential rate and fix the ratios analytically.  Since the
coefficients are polynomial in $\nu$, a single evaluation of the conical
function at each $(\tau,\lambda)$ serves every block at once, and that is
what keeps the numerics of Appendix~\ref{app:valid} stable.

\section{Wigner-positive pure states beyond the coherent orbit}
\label{sec:beyond}

The lowest weight state $\ket{k,k}$ is a coherent state, and $W_{00}$ is
strictly positive everywhere.  The first excited state is not a coherent
state, and $W_{11}$ takes negative values on a neighbourhood of the
origin, reaching $-4.01$ there at $k=1$.  A superposition of the two sets
a positive profile against a negative one, and the question is how much
of the second the first can absorb.

Consider the two-state family
\begin{equation}
  \ket{\psi_\theta}=\cos\theta\,\ket{k,k}+\sin\theta\,\ket{k,k+1}.
  \label{eq:family}
\end{equation}
Table~\ref{tab:blocks} gives its Wigner function as
\begin{align}
  W_\theta(\tau,\phi)
  &=\cos^2\!\theta\,W_{00}(\tau)+\sin^2\!\theta\,W_{11}(\tau)
  \nonumber\\
  &\quad -2\sin\theta\cos\theta\,C(\tau)\cos\phi ,
  \label{eq:Wtheta}
\end{align}
where the population of the excited level contributes
\begin{equation}
  W_{11}(\tau)=\frac{1}{(2k-1)k}\int_0^\infty \dOm \;
  \bigl(2k+\nu\bigr)\,\Pc(\cosh\tau)
  \label{eq:W11}
\end{equation}
and the coherence between the two levels contributes $C$, the radial part
of the $(0,1)$ block, fixed by $\mathrm{Re}\,W_{01}=-C(\tau)\cos\phi$ and
given by
\begin{equation}
  C(\tau)=\frac{2\sinh\tau}{\sqrt{2k}\,(2k-1)}
  \int_0^\infty \dOm \; \Pc'(\cosh\tau).
  \label{eq:C}
\end{equation}
Both follow from Table~\ref{tab:blocks} on restoring the normalisation
$N_{k;mn}$ of Eq.~\eqref{eq:Nmn}.

Before evaluating this on the hyperboloid it pays to run the same family
through the flat case, where the answer is known and sharp.  For a single
mode the Fock blocks are elementary~\cite{CahillGlauber1969}, and
$\cos\theta\ket{0}+\sin\theta\ket{1}$ has the Wigner function
$(2/\pi)e^{-2|\alpha|^2}f(|\alpha|)$ with
\begin{equation}
  f(t)=\cos2\theta+4\sin^2\!\theta\,t^2-4\sin\theta\cos\theta\,t .
  \label{eq:flat}
\end{equation}
The discriminant of Eq.~\eqref{eq:flat} is $16\sin^4\!\theta$, which is
positive for every $\theta\ne0$, so $f$ has real roots and the Wigner
function goes negative.  The minimum value is $-\sin^2\!\theta$ and it
sits at $|\alpha|=\cot\theta/2$.  Hudson's theorem appears here in a
concrete form.  The admissible window is exactly a point, the negativity
retreats to infinity as $\theta\to0$ rather than disappearing, and it
shrinks only quadratically in the mixing angle.

The runaway is the mechanism worth keeping in mind.  In the flat case the
two ratios $W_{11}/W_{00}=4|\alpha|^2-1$ and $|C|/W_{00}=2|\alpha|$ both
diverge, so the band of forbidden angles slides towards zero and however
small $\theta$ is there is always a radius far enough out to catch it.
Section~\ref{sec:kdep} shows the hyperboloid closes that escape route.

Minimising Eq.~\eqref{eq:Wtheta} over $\phi$ is immediate.  Giving the
excited amplitude a relative phase $e^{i\delta}$ replaces $\cos\phi$ by
$\cos(\phi+\delta)$ and so changes nothing after that minimisation.
Hence $W_\theta\ge0$ on the whole hyperboloid if and only if
\begin{align}
  \cos^2\!\theta\,W_{00}(\tau)
  &+\sin^2\!\theta\,W_{11}(\tau)
  \nonumber\\
  &-2|\sin\theta\cos\theta|\,|C(\tau)| \;\ge\; 0
  \label{eq:criterion}
\end{align}
for every $\tau\ge0$.  Positivity is therefore a one-parameter question
at each $\tau$, and the admissible $\theta$ is bounded by the smaller
root of a quadratic.

\begin{figure}[t]
\includegraphics[width=\columnwidth]{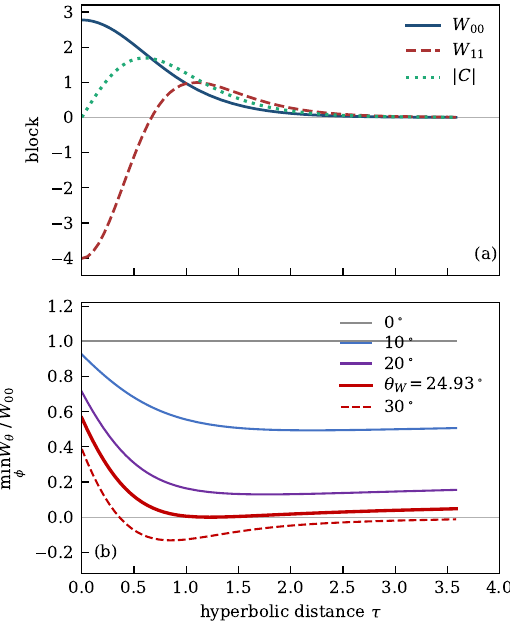}
\caption{\label{fig:one}(a) The three radial blocks at $k=1$.  $W_{00}$
stays positive, $W_{11}$ is negative out to $\tau\simeq0.67$, and the
coherence $|C|$ peaks at $\tau\simeq0.60$.  (b) The positivity criterion
\eqref{eq:criterion}, normalised by $W_{00}$, for five mixing angles.
The curve at $\theta_W$ touches zero at $\tau=1.225$, away
from both the origin and the tail, and the curve at $30^\circ$ dips
below.  Normalising by $W_{00}$ removes the common exponential decay and
makes the margin quoted in the text visible.}
\end{figure}

The three functions play distinct roles, and Fig.~\ref{fig:one}(a) shows
them.  $W_{00}$ is positive everywhere and supplies the budget.  $W_{11}$
changes sign and is negative near the origin.  The coherence $C$ enters
through $\cos\phi$, so whatever its sign it subtracts somewhere on each
orbit of the rotation generated by $\hat K_0$.

Evaluating Eq.~\eqref{eq:criterion} at $k=1$ gives
\begin{equation}
  \theta_W=24.92^\circ ,
  \qquad
  \sin^2\!\theta_W=0.178 ,
  \label{eq:thetaW}
\end{equation}
as the largest admissible mixing angle.  The Wigner-positive pure states
contain the Perelomov coherent orbit strictly, and Hudson's dichotomy has
no analogue here.

Comparing with Eq.~\eqref{eq:flat} makes the content of
Eq.~\eqref{eq:thetaW} plain.  The flat family is negative for every
$\theta>0$ and the hyperbolic family is positive up to a finite angle, so
the two definitions of classicality that Hudson's theorem identifies come
apart as soon as the phase space is curved.  The gap is not a boundary
effect either.  At $\theta=20^\circ$ the state carries $12\%$ of its
weight on the first excited level and its Wigner function stays positive
across the whole hyperboloid.

The binding value of $\tau$ is $1.225$ at $k=1$
[Fig.~\ref{fig:one}(b)].  It sits away from the origin, where $W_{11}$ is
deepest but $W_{00}$ is also largest, and away from the tail, where
Sec.~\ref{sec:kdep} shows the quadratic form becomes marginal.  The
threshold is therefore fixed at intermediate hyperbolic distance, and the
margin elsewhere stays finite.

The positivity is uniform rather than marginal in the exponentially small
tail, where quadrature noise would otherwise decide the answer.  The
ratio $\min_\tau W_\theta/W_{00}$ equals $0.129$ at $\theta=20^\circ$ and
$0.023$ at $\theta=24^\circ$, and it does not decrease as $\tau$ grows.

That margin is what lets us pass from a curve of states to a set.  The
symbol $W_A$ depends linearly on $A$, and on the region where the margin
is bounded below the dependence is uniform, so a small enough
trace-norm perturbation cannot change the sign and an open ball of states
around $\ket{\psi_\theta}$ stays inside the positive set.  Adding the coset directions of $SU(1,1)/U(1)$
supplies two further parameters, so the Wigner-positive pure states form
a set of positive volume, of dimension at least three against the two of
the coherent orbit.

The argument just given rests on continuity, and it leaves open whether
the enlargement is confined to a thin neighbourhood of the single
direction we chose.  We therefore repeat the calculation with a
third level, which tests both the size of the admissible set and the
special status of the two-state direction.

Direct computation on the three-state family
$c_0\ket{k,k}+c_1\ket{k,k+1}+c_2\ket{k,k+2}$ sharpens the picture.
Writing $c_0=\cos\chi$, $c_1=\sin\chi\cos\beta$ and
$c_2=\sin\chi\sin\beta$, then optimising over the single free relative
phase, the admissible $\chi$ is largest away from the two-state
direction (Table~\ref{tab:three}).  The maximum sits at
$\beta\simeq20^\circ$, where $\chi$ reaches $39.05^\circ$ against
$24.92^\circ$ at $\beta=0$, the latter reproducing
Eq.~\eqref{eq:thetaW}.  Sampling $\chi$ from $0^\circ$ to $45^\circ$ in steps of $1^\circ$ and
$\beta$ from $0^\circ$ to $90^\circ$ in steps of $5^\circ$, and optimising
the relative phase at each point, the admissible region covers $55\%$ of
that rectangle.

\begin{table}[tb]
\caption{\label{tab:three}Largest admissible $\chi$ for the three-state
family at $k=1$, against the mixing angle $\beta$ that controls the
weight on $\ket{k,k+2}$.  Entries are located by bisection in
$\chi$ and quoted to the accuracy set by the discretisation of $\tau$,
$\phi$ and the relative phase, which we estimate at a few times
$10^{-2}$ degrees.  The entry at $\beta=0$
reproduces Eq.~\eqref{eq:thetaW}.}
\begin{ruledtabular}
\begin{tabular}{cc}
$\beta$ (deg) & $\chi_{\max}$ (deg) \\
\hline
0  & 24.92 \\
5  & 28.24 \\
10 & 32.63 \\
15 & 38.11 \\
20 & 39.05 \\
25 & 34.76 \\
30 & 30.81 \\
40 & 25.15 \\
60 & 19.32 \\
90 & 16.92 \\
\end{tabular}
\end{ruledtabular}
\end{table}

The widening has a structural source.  The $(0,2)$ coherence enters
through $\cos(2\phi+\psi)$ while the $(0,1)$ and $(1,2)$ coherences enter
through the first harmonic, so the two harmonics can be phased against
each other to reduce the deepest excursion in $\phi$.  Adding a little
$\ket{k,k+2}$ therefore buys back some of the coherence penalty that
Eq.~\eqref{eq:criterion} charges.  The admissible set bulges in that
direction, and a group orbit carries no such feature, which is evidence
against any purely group-theoretic characterisation of the
Wigner-positive class.

What the section establishes is that the Wigner-positive pure states on
the hyperboloid form a set of positive volume that properly contains the
coherent orbit, with a threshold at $k=1$ that two independent
implementations agree on to $6\times10^{-4}$ degrees.  What it does not
establish is any structural description of
that set.  We know its dimension exceeds the orbit's, we know it has a
boundary, and we know the boundary is not attained in the direction one
would guess.  The next section explains why the threshold sits where it
does.

\section{Asymptotic origin of the window}
\label{sec:kdep}

Equation~\eqref{eq:thetaW} fixes the threshold at $k=1$.  That it is
nonzero at all is the surprise, since the flat case leaves no room, and
the large-$k$ limit must recover the flat case.  The In\"on\"u-Wigner contraction of $SU(1,1)$ to the Heisenberg-Weyl
algebra~\cite{Perelomov1986,Kam2023} sends $\hat K_-/\sqrt{2k}$ to the
annihilation operator and the Perelomov coherent states to Gaussian
coherent states, so $\theta_W$ has to close as $k$ grows.  The question is what keeps it
open at finite $k$, and the answer lies in the far field.

Write $R=W_{11}/W_{00}$ and $S=|C|/W_{00}$, and call the region of large
hyperbolic distance the far field.  Both ratios approach finite limits
there,
\begin{equation}
  R\to 2k ,
  \qquad
  S\to\sqrt{2k} ,
  \label{eq:RS}
\end{equation}
as $\tau\to\infty$, with the approach tabulated in
Appendix~\ref{app:valid}, so that the discriminant of the quadratic in
Eq.~\eqref{eq:criterion} obeys
\begin{equation}
  S^2-R\;\to\;0 .
  \label{eq:critical}
\end{equation}
The three functions share one branch point of $\Phi_k^{1/2}$, the one
nearest the real $\lambda$ axis at $|\mathrm{Im}\,\lambda|=2k-\frac12$,
which fixes a common exponential rate and leaves only the polynomial and
derivative factors to distinguish them.  Appendix~\ref{app:asym} carries
out the contour argument and Appendix~\ref{app:valid} reports the
numerical confirmation.

Equation~\eqref{eq:critical} says the $2\times2$ form built from
$W_{00}$, $W_{11}$ and $C$ becomes rank deficient in the far field.  Its
null direction is fixed by $S^2=R$, which turns the quadratic into a
perfect square and gives
\begin{equation}
  \tan\theta_{\rm null}=\frac{1}{\sqrt{R}}=\frac{1}{\sqrt{2k}} .
  \label{eq:null}
\end{equation}
The degeneracy is approached rather than reached, and slowly.  The
discriminant stays small and positive at finite $\tau$, falling as
\begin{equation}
  S^2-R \;\simeq\; \frac{1}{4k\tau} ,
  \label{eq:rate2}
\end{equation}
which we obtain from the numerical values at $k=1$ and $k=2$.  The far
field therefore forbids a narrow band of mixing angles rather than
none.  At $k=1$ the
band runs from $27.7^\circ$ to $38.2^\circ$ at $\tau=3$ and from
$32.6^\circ$ to $37.3^\circ$ at $\tau=16$, closing onto
$\theta_{\rm null}=35.26^\circ$.  It never reaches $\theta_W=24.9^\circ$.

Equation~\eqref{eq:null} has a second reading.  Expanding
Eq.~\eqref{eq:coh} for the family~\eqref{eq:family} gives the Husimi
amplitude
\begin{equation}
  \braket{\zeta}{\psi_\theta}\propto(1-|\zeta|^2)^k
  \bigl[\cos\theta+\sqrt{2k}\,\sin\theta\;\bar\zeta\bigr],
  \label{eq:husamp}
\end{equation}
whose single zero sits at $|\zeta|=\cot\theta/\sqrt{2k}$ and crosses the
boundary of the Poincar\'e disc exactly when
$\tan\theta=1/\sqrt{2k}$.  The stellar rank counts the zeros of the Husimi amplitude, and it
classifies states by how far they sit from the coherent
orbit~\cite{Chabaud2020}.  The angle at which the far field degenerates
is therefore the angle at which the state acquires its first Husimi zero,
which is where the stellar rank steps from zero to one.  The
degeneracy of the quadratic form and the entry of a zero into the disc
are the same event.

The flat case fails for want of this closure.  Repeating the
construction on Eq.~\eqref{eq:flat} gives $R=4|\alpha|^2-1$ and
$S=2|\alpha|$, so the discriminant $S^2-R$ equals unity identically while
both ratios diverge.  The forbidden band therefore slides downwards
rather than narrowing onto a fixed angle.  Its edges are
$\tan\theta=1/(2|\alpha|\pm1)$, which run from $18.4^\circ$ to
$45.0^\circ$ at $|\alpha|=1$ and from $0.94^\circ$ to $0.97^\circ$ at
$|\alpha|=30$.  Every mixing angle is eventually caught, which is
Hudson's theorem seen in the far field.

On the hyperboloid the band stalls at a fixed nonzero angle, so every
$\theta$ below $\theta_{\rm null}$ escapes the far field entirely and the
binding constraint has to come from intermediate $\tau$.  This is what
Fig.~\ref{fig:one}(b) shows, with the deciding minimum at $\tau=1.225$.
The window is finite because the geometry is curved, and it closes as
$k\to\infty$ because Eq.~\eqref{eq:null} drives $\theta_{\rm null}$ to
zero.

The same feature sets the practical limit on what we can report.  Because
the deciding region is a plateau at intermediate $\tau$ rather than a
sharp minimum, the numerical threshold at larger $k$ depends on where the
exponentially decaying tail is truncated, and we have not found a
truncation-independent value beyond $k$ of order a few.  We therefore
quote $\theta_W$ at $k=1$, where the minimum is sharp and the value is
stable across four decades of truncation level, and we leave the $k$
dependence open.

\section{Discussion}
\label{sec:disc}

Hudson's theorem identifies two notions of classicality, one supplied by
the symmetry and one supplied by the phase-space function.  On the
hyperboloid they come apart.  The Wigner-positive pure states contain the
Perelomov coherent orbit strictly, they outnumber it by at least one
parameter, and the excess is not small.  A state carrying $12\%$ of its
weight on the first excited level still has a Wigner function that is
positive everywhere, and the admissible mixing angle reaches
$24.9^\circ$ at $k=1$.

The mechanism is a curvature effect and it can be stated in one line.
Positivity is decided by a competition between the coherence and the two
populations, and in the flat case that competition is lost at large
distance because both ratios $W_{11}/W_{00}$ and $|C|/W_{00}$ diverge
there.  On the hyperboloid the same ratios saturate, at $2k$ and
$\sqrt{2k}$, because all three blocks are governed by one branch point of
$\Phi_k^{1/2}$ whose position depends on the representation label rather
than on the phase-space point.  Flat phase space carries no such label,
and the corresponding rate is set by $|\alpha|$ itself, which is why
nothing there stops the ratios from growing.  The forbidden band therefore stalls at
$\theta_{\rm null}=\arctan(1/\sqrt{2k})$ instead of sweeping down to
zero, and the angles below it survive.  Equation~\eqref{eq:null} also
identifies that angle with the point at which a Husimi zero enters the
disc, so the far-field degeneracy and the first step of the stellar
hierarchy are the same event.

The practical consequence is a warning about transfer.  Wigner negativity
is used on flat phase space as an indicator of
non-classicality~\cite{Kenfack2004}, as a bound on classical
simulability~\cite{MariEisert2012}, and as the quantity equivalent to
contextuality for continuous-variable
measurements~\cite{BoothChabaudEmeriau2022}.  Each of these readings
borrows the identification that Hudson's theorem supplies.  On the
hyperboloid the identification is unavailable, so a reconstructed
$SU(1,1)$ Wigner function that comes out non-negative certifies neither a
coherent state nor a vanishing stellar rank, and arguments of the
simulability type would have to be rebuilt on whatever free class the
geometry actually singles out.

Several limits of the present treatment are worth stating plainly.  We
work with pure states in the positive discrete series at $k\ge1$ and use
only the three lowest levels, and we characterise the Wigner-positive set
by exhibiting points of it rather than by describing it.  Its boundary is
attained away from the direction one would guess, which already argues
against a purely group-theoretic description.  Mixed states are untouched
here, and even in the flat case the corresponding statement is only
partly settled~\cite{Mandilara2009}.

Two directions follow.  The contraction of $SU(1,1)$ to the
Heisenberg-Weyl algebra turns $k$ into a controlled interpolation between
the two cases treated above, and tracking $\theta_W$ along it would give
the rate at which Hudson's theorem is restored, which is the $k$
dependence we could not stabilise numerically.  Second, the sphere and
the hyperboloid fail in opposite ways, the first for want of a free class
and the second for having too large a one, so a proof of the conjecture
that no pure spin state has a non-negative $SU(2)$ Wigner
function~\cite{Davis2021,Davis2023} would complete the picture for
constant curvature.

\section*{Data availability}

The code that reproduces the numbers, the tables and the figure of this
paper is archived at
\href{https://doi.org/10.5281/zenodo.22310147}{10.5281/zenodo.22310147},
together with the specification and the output of the second
implementation described in Appendix~\ref{app:valid}.

\appendix

\section{Reduction to two radial functions}
\label{app:lemma}

Klimov \emph{et al.} give the $P$ symbol of
$\ket{k,k+m}\bra{k,k+n}$ as a derivative of the $P$ symbol of the lowest
weight state, and, after substitution into the $P$ to $W$ convolution and
one integration by parts, obtain \cite[Eq.~(B.6)]{Klimov2020}
\begin{align}
  W_{mn}(\zeta)=\ &\frac{N_{k;mn}}{(2k-1)\pi}\int_0^\infty \dOm
  \nonumber\\[2pt]
  &\times\int d\tau'd\phi'\,\delta(\tau')\,
  \partial_{\zeta'}^m\partial_{\zeta'^*}^n\,\mathcal{G} ,
  \label{eq:B6}
\end{align}
where
\begin{equation}
  \mathcal{G}=\cosh^{4k}(\tau'/2)\,\Pc(\cosh\xi),
  \label{eq:Gdef}
\end{equation}
\begin{equation}
  N_{k;mn}=\frac{\Gamma(2k)}
  {\sqrt{m!\,n!\,\Gamma(m{+}2k)\,\Gamma(n{+}2k)}},
  \label{eq:Nmn}
\end{equation}
and
\begin{equation}
  \cosh\xi=\cosh\tau\cosh\tau'
  -\cos(\phi-\phi')\sinh\tau\sinh\tau' .
  \label{eq:coshxi}
\end{equation}

The derivatives in Eq.~\eqref{eq:B6} are ordinary derivatives in the
complex disc coordinate $\zeta'$.  Reference~\cite{Klimov2020} writes
them out in polar coordinates, where they carry a factor
$\coth(\tau'/2)$ that diverges at the origin, but that divergence is the
usual artefact of polar coordinates at their centre.  In the complex
coordinate the operator is regular there, and since $\delta(\tau')$ sets
$\zeta'=0$ the inner object of Eq.~\eqref{eq:B6} is simply $m!\,n!$ times
a Taylor coefficient at the origin.  That coefficient depends on $\zeta$
alone, so the remaining $\phi'$ integration contributes a factor $2\pi$
and the prefactor of Eq.~\eqref{eq:B6} collapses to
$2N_{k;mn}/(2k-1)$.

Write $z=\zeta'$.  With $\cosh^{4k}(\tau'/2)=(1-|z|^2)^{-2k}$ and the
disc form of the pseudo-scalar product~\cite[Eq.~(B.11)]{Klimov2020},
\begin{equation}
  \cosh\xi=\frac{2|1-\zeta^*z|^2}{(1-|\zeta|^2)(1-|z|^2)}-1 ,
  \label{eq:B11}
\end{equation}
and using $u_0+1=2/(1-|\zeta|^2)$ with $u_0=\cosh\tau$, the argument of
the conical function becomes
\begin{equation}
  u(z,\bar z)+1
  =\frac{(u_0+1)(1-\zeta^*z)(1-\zeta\bar z)}{1-z\bar z} .
  \label{eq:ushift}
\end{equation}
The generating function is therefore
\begin{equation}
  F(z,\bar z)=(1-z\bar z)^{-2k}\,\Pc\bigl(u(z,\bar z)\bigr),
  \label{eq:genfun}
\end{equation}
and $W_{mn}$ is proportional to
$\partial_z^m\partial_{\bar z}^n F\bigl|_0$.

Expanding Eq.~\eqref{eq:ushift} needs the geometric series
$(1-z\bar z)^{-1}=\sum_{l\ge0}(z\bar z)^l$.  Multiplying it by
$(1-\zeta^*z)(1-\zeta\bar z)$ and collecting terms, the coefficient of
$z\bar z$ in $u+1$ is $(u_0+1)(1+|\zeta|^2)$, and
$|\zeta|^2=(u_0-1)/(u_0+1)$ turns that into $2u_0$.  Hence
\begin{equation}
  u-u_0=a\,z+\bar a\,\bar z+2u_0\,z\bar z+O(3) ,
  \qquad a=-(u_0+1)\zeta^* ,
  \label{eq:duexp}
\end{equation}
and $|a|=(u_0+1)|\zeta|=\sinh\tau$.  Since
$\zeta=\tanh(\tau/2)e^{-i\phi}$, each power of $\bar z$ brings one factor
$\bar a\propto\zeta$ and therefore one factor
$\tanh(\tau/2)e^{-i\phi}$, while each power of $z$ brings the conjugate.
This fixes the angular factor $\tanh^{|m-n|}(\tau/2)e^{-i(n-m)\phi}$
stripped in Table~\ref{tab:blocks}.

Composing the Taylor series of $\Pc$ about $u_0$ with
Eq.~\eqref{eq:duexp} and with the binomial series of $(1-z\bar z)^{-2k}$
produces derivatives $\Pc^{(j)}(u_0)$ up to $j=m+n$.  Differentiating
Eq.~\eqref{eq:legendre} $j$ times gives
\begin{align}
  (u^2-1)P^{(j+2)}&+2(j+1)\,u\,P^{(j+1)}
  \nonumber\\
  &+\bigl[j(j+1)-\nu\bigr]P^{(j)}=0 ,
  \label{eq:legrec}
\end{align}
which expresses every derivative above the first through $\Pc$ and
$\Pc'$ with coefficients rational in $u$.  Collecting terms gives
Table~\ref{tab:blocks}.  We carried the collection out symbolically and
checked it against an independent numerical expansion, as
Appendix~\ref{app:valid} records.

The diagonal half of the selection rule has a structural origin.
Equation~(3.11) of Ref.~\cite{Klimov2020} generates the block with
$m=n$ by $m$ applications of the Laplacian on the hyperboloid, and the
conical function is an eigenfunction of that operator with eigenvalue
$\nu$, so the coefficient is a polynomial of degree $m$ in $\nu$
multiplying $\Pc$ with no $\Pc'$ left over.  The cancellation is visible
already at $m=1$, where the term $2u_0\Pc'$ coming from
Eq.~\eqref{eq:duexp} is removed exactly by the $-2u\Pc'$ that
Eq.~\eqref{eq:legrec} produces at $j=0$.  The off-diagonal half of the
rule, that $|m-n|=1$ leaves only $\Pc'$ and that $|m-n|\ge2$ mixes the
two, we read off from the computed coefficients rather than prove in
general.

As a check, setting $m=n=1$ gives the coefficient
$2k+\nu=2k-\frac14-\lambda^2$, which reproduces
Ref.~\cite[Eq.~(3.12)]{Klimov2020}.  Appendix~\ref{app:valid} records the
further numerical checks.

\section{Asymptotics of the ratios}
\label{app:asym}

All three functions of Sec.~\ref{sec:beyond} are integrals of the
conical function against the common measure $\dOm$ of
Eq.~\eqref{eq:dOm}.  Their large-$\tau$ behaviour therefore follows from
one analysis.

For large argument the Legendre function of the first kind
obeys~\cite{Erdelyi1955,DLMF}
\begin{equation}
  P_\mu(u)\sim
  \frac{\Gamma(\mu+\tfrac12)}{\sqrt\pi\,\Gamma(\mu+1)}(2u)^{\mu}
  +\frac{\Gamma(-\mu-\tfrac12)}{\sqrt\pi\,\Gamma(-\mu)}(2u)^{-\mu-1} .
  \label{eq:Pgen}
\end{equation}
Setting $\mu=-\tfrac12+i\lambda$ makes the second term the complex
conjugate of the first, so
\begin{equation}
  \Pc(u)\;\sim\;
  \frac{\Gamma(i\lambda)}{\sqrt{\pi}\,\Gamma(\tfrac12+i\lambda)}
  \,(2u)^{-\frac12+i\lambda}
  \;+\;\text{c.c.} ,
  \label{eq:Pasym}
\end{equation}
and $(2u)^{-1/2+i\lambda}\sim e^{(-\frac12+i\lambda)\tau}$ at large
$\tau$.

Before deforming anything we extend the range.  The measure
$\dOm$ is even in $\lambda$, because $\lambda\tanh(\pi\lambda)$ and
$\Phi_k$ both are, and the conical function satisfies
$\Pc=P_{-1/2-i\lambda}$ by the symmetry $P_\mu=P_{-\mu-1}$, so the whole
integrand is even and
$\int_0^\infty=\tfrac12\int_{-\infty}^{\infty}$.  The point $\lambda=0$
is regular, since $\lambda\tanh(\pi\lambda)$ vanishes quadratically
there.  On the full line the branch of Eq.~\eqref{eq:Pasym} carrying
$e^{i\lambda\tau}$ is closed in the upper half plane and its conjugate in
the lower, and the rate is set by the singularity of the integrand
nearest the real axis.

Two families of singularities compete, and one of them drops out.  The
factor $\tanh(\pi\lambda)$ has simple poles at
$\lambda=i(n+\tfrac12)$, $n\ge0$, the nearest at $\lambda=i/2$.  At those
points $\Gamma(\tfrac12+i\lambda)=\Gamma(-n)$ diverges, so the amplitude
in Eq.~\eqref{eq:Pasym} carries a simple zero there.  The poles of
$\tanh(\pi\lambda)$ are cancelled, and they contribute nothing.

What remains is $\Phi_k^{1/2}$.  By Eq.~\eqref{eq:Phi} its square is
built from $\Gamma(2k-\tfrac12+i\lambda)\Gamma(2k-\tfrac12-i\lambda)$,
whose poles sit at $\lambda=\pm i(2k-\tfrac12+n)$ with $n\ge0$.  Taking
the square root turns these into branch points, and the nearest lies at
\begin{equation}
  |\mathrm{Im}\,\lambda_*|=2k-\tfrac12 .
  \label{eq:branch}
\end{equation}
The singularity at $\lambda_*$ is a branch point rather than a pole, so
deforming the contour wraps a cut rather than picking up a residue.  The
cut contributes $e^{i\lambda_*\tau}=e^{-(2k-1/2)\tau}$ together with an
algebraic prefactor in $\tau$, and the prefactor $e^{-\tau/2}$ of
Eq.~\eqref{eq:Pasym} completes the exponential rate,
\begin{equation}
  W_{00},\;W_{11},\;C \;\sim\; e^{-2k\tau} .
  \label{eq:rate}
\end{equation}
The three functions wrap the same cut, so the algebraic prefactor is
common to them and cancels in any ratio, along with the exponential.
What distinguishes them is the factor each carries under the integral,
and those factors are analytic and slowly varying across the cut, so to
leading order they may be taken outside at $\lambda_*$.

For $W_{11}$ the extra factor is $2k-\tfrac14-\lambda^2$.  At
$\lambda_*^2=-(2k-\tfrac12)^2$ it equals
\begin{equation}
  2k-\tfrac14+\bigl(2k-\tfrac12\bigr)^2=4k^2 ,
\end{equation}
so, restoring the prefactors of Eqs.~\eqref{eq:W00} and
\eqref{eq:Wtheta},
\begin{equation}
  R=\frac{W_{11}}{W_{00}}
  \;\longrightarrow\;
  \frac{4k^2}{(2k-1)k}\cdot\frac{2k-1}{2}=2k .
  \label{eq:Rlim}
\end{equation}

For $C$ the extra factor is a derivative.  Since $u=\cosh\tau$ we have
$\Pc'=(d\Pc/d\tau)/\sinh\tau$, and Eq.~\eqref{eq:Pasym} gives
$d\Pc/d\tau\to(-\tfrac12+i\lambda)\Pc$.  At $\lambda_*$ this factor is
\begin{equation}
  -\tfrac12-\bigl(2k-\tfrac12\bigr)=-2k .
\end{equation}
The explicit $\sinh\tau$ in Eq.~\eqref{eq:C} cancels the one from the
derivative, and
\begin{equation}
  S=\frac{|C|}{W_{00}}
  \;\longrightarrow\;
  \frac{4k}{\sqrt{2k}\,(2k-1)}\cdot\frac{2k-1}{2}=\sqrt{2k} .
  \label{eq:Slim}
\end{equation}
Equations~\eqref{eq:Rlim} and \eqref{eq:Slim} give $S^2-R\to0$, which is Eq.~\eqref{eq:critical}.

The argument fixes the rate and the two ratios, which is all
Sec.~\ref{sec:kdep} needs, and it leaves the overall prefactor
undetermined.  The second branch of Eq.~\eqref{eq:Pasym} repeats the
same computation with $\lambda_*$ conjugated and reproduces the same
ratios, so the real combination inherits them.
Appendix~\ref{app:valid} reports the numerical confirmation of
Eq.~\eqref{eq:Rlim}.

\section{Validation and numerics}
\label{app:valid}

Self-duality gives a family of identities that every block must satisfy,
and we use them as the main check on the whole construction.  Writing
\begin{equation}
  I[f]=\frac{2k-1}{2}\int f(\tau)\,\sinh\tau\,d\tau ,
  \label{eq:Ifunctional}
\end{equation}
Eq.~\eqref{eq:selfdual} requires $I[\,|W_{mn}|^2]=1$ for every block and
$I[W_{mm}W_{nn}]=0$ whenever $m\ne n$.  At $k=1$ the three diagonal
blocks with $m=0,1,2$ return $1.0002$, $0.9979$ and $1.0003$.  The blocks
$(0,1)$, $(0,2)$ and $(1,2)$ return $1.0011$, $1.0005$ and $0.9994$.  The
three cross terms return $0.00133$, $-0.00354$ and $0.00127$.  These
identities involve the normalisations, the angular factors and the
$\lambda$ integration at once, so passing them at the level of a few
parts in $10^4$ constrains all three together.

The coefficients of Table~\ref{tab:blocks} were obtained twice.  We
differentiated Eq.~\eqref{eq:genfun} symbolically and, independently,
expanded it as a numerical bivariate power series with the derivatives of
the conical function generated by Eq.~\eqref{eq:legrec}.  The two agree
to machine precision on every block up to $(3,3)$, and the entry with
$m=n=1$ agrees with Eq.~(3.12) of Ref.~\cite{Klimov2020}.

A second implementation, built on different representations of the
conical function and a different quadrature rule, reproduces these
results at higher precision.  It satisfies the four identities
$I[W_{00}^2]=I[W_{11}^2]=I[C^2]=1$ and $I[W_{00}W_{11}]=0$ with residuals
below $10^{-13}$, returns $\theta_W=24.9203^\circ$ with binding
$\tau=1.2251$, and gives the same threshold to six decimals at every
truncation level between $10^{-4}$ and $10^{-8}$.  The two
implementations agree on $\theta_W$ to $6\times10^{-4}$ degrees.  What
limits the residuals in either implementation is the quadrature in
$\tau$ rather than the one in $\lambda$.

The asymptotic ratios of Sec.~\ref{sec:kdep} are approached from above,
and Table~\ref{tab:asym} lists them together with the combination that
Eq.~\eqref{eq:rate2} predicts to tend to $1/(4k)$.  We also checked the strict positivity of $W_{00}$ that
Ref.~\cite{Klimov2026} proves.  How far out the check can be pushed is
set by cancellation in the $\lambda$ integral, and two settings control
it.  We therefore accept a distance only when the value is stable both
under a change of the segmentation and under a change of the upper
cutoff, to six digits in each case.  The second condition is the
binding one.  At $k=2$ and $\xi=20$ the cutoff of
Eq.~\eqref{eq:cutoff}, which serves everywhere else in this paper,
truncates the integral where the tail still dominates.  The value it
returns exceeds the converged one by more than an order of magnitude,
and for nearby choices of $\Lambda$ it comes out negative, while two
segmentations of that truncated integral still agree to eight digits.  For the deep-distance
check we take $\Lambda=60$ instead, beyond which the value no longer
moves up to $\Lambda=110$.

On that criterion the check reaches $\xi=20$ at $k=1$ and $k=2$,
$\xi=6$ at $k=5$, and $\xi=3$ at $k=10$, with no change of sign
anywhere in those ranges.  At $k=1$ the values fall from $2.7761$ at the
origin to $6.249\times10^{-18}$ at $\xi=20$, and at $k=2$ they reach
$7.806\times10^{-35}$ at the same distance.

\begin{table}[t]
\caption{\label{tab:asym}The two ratios of Sec.~\ref{sec:kdep}, whose
limits are $R\to2k$ and $S\to\sqrt{2k}$.  The last column, which
Eq.~\eqref{eq:rate2} sends to $1/(4k)$, is computed from unrounded data
and approaches $0.25$ and $0.125$ respectively.  The second block stops at $\tau=8$ because
the ratios need $W_{11}$ and $C$ as well, whose integrands carry the
extra factor $2k-\frac14-\lambda^2$ and lose relative accuracy well
before $W_{00}$ does.}
\begin{ruledtabular}
\begin{tabular}{ccccc}
$k$ & $\tau$ & $R$ & $S$ & $\tau(S^2-R)$ \\
\hline
1 & 3  & 2.4193 & 1.5873 & 0.3004 \\
1 & 5  & 2.2409 & 1.5143 & 0.2611 \\
1 & 8  & 2.1227 & 1.4679 & 0.2552 \\
1 & 12 & 2.0733 & 1.4472 & 0.2532 \\
\hline
2 & 3  & 4.4714 & 2.1285 & 0.1776 \\
2 & 5  & 4.2820 & 2.0760 & 0.1386 \\
2 & 8  & 4.1434 & 2.0396 & 0.1320 \\
\end{tabular}
\end{ruledtabular}
\end{table}

Conical functions come from \texttt{mpmath}.  We obtain the derivative
from the exact identity
\begin{equation}
  (u^2-1)P'_\mu(u)=\mu\bigl[uP_\mu(u)-P_{\mu-1}(u)\bigr],
  \qquad \mu=-\tfrac12+i\lambda ,
  \label{eq:dP}
\end{equation}
which avoids finite differences, and we integrate over $\lambda$ by
Gauss-Legendre quadrature with the cutoff
\begin{equation}
  \Lambda=\frac{4k}{\pi}+45 ,
  \label{eq:cutoff}
\end{equation}
since the support of $\Phi_k^{1/2}$ moves outwards with $k$.  This rule
is adequate at the hyperbolic distances that decide the threshold and
fails at the much larger distances used for the positivity check above,
where the integrand oscillates fast enough that the truncated tail still
matters.  Two further points are worth recording for anyone repeating
the calculation.  The quadrature nodes have to be generated at the
working precision, since nodes imported at double precision impose an
error floor that no increase of the working precision removes.  And the
ratios $R$ and $S$ lose relative accuracy before $W_{00}$ does, for the
reason given in the caption of Table~\ref{tab:asym}.  Threshold
detection uses the ratio to $W_{00}$ rather than the absolute value.  The
reason is that $W$ decays exponentially in $\tau$, so an absolute
criterion applied in the tail tests quadrature noise instead of the
function.

Two limitations deserve to be stated plainly.  The normalisation checks
hold to a few parts in $10^4$ at $k=1$ and drift to about $10^{-3}$ by
$k=16$.  The threshold is more delicate than that figure suggests.  At
$k=1$ the deciding minimum in $\tau$ is sharp, and $\theta_W$ is
unchanged to six decimals when the tail truncation is varied over four
decades, from $10^{-4}$ to $10^{-8}$ of the peak of $W_{00}$.  At larger $k$ the deciding region flattens into a
plateau near $\tau\simeq1.5$, the truncation level moves the cut across
that plateau, and the extracted threshold shifts by several tenths of a
degree.  Raising the quadrature density in $\lambda$ by an order of
magnitude improves the normalisation checks while leaving that shift
unchanged, which identifies the cause as the flatness of the deciding
region rather than quadrature error.  We therefore report the threshold
only at $k=1$.

The second limitation is one of range.  Everything above uses blocks with
$m,n\le2$.  Higher blocks follow from Table~\ref{tab:blocks}, but their
numerical evaluation is not stable at the precision used here, and no
result in this paper relies on them.

\bibliography{refs}

@article{Groenewold1946,
  title = {On the principles of elementary quantum mechanics},
  author = {Groenewold, H. J.}, journal = {Physica}, volume = {12},
  pages = {405}, year = {1946}}

@article{Moyal1949,
  title = {Quantum mechanics as a statistical theory},
  author = {Moyal, J. E.}, journal = {Proc. Cambridge Philos. Soc.},
  volume = {45}, pages = {99}, year = {1949}}

@article{Hillery1984,
  title = {Distribution functions in physics: Fundamentals},
  author = {Hillery, M. and O'Connell, R. F. and Scully, M. O. and Wigner, E. P.},
  journal = {Phys. Rep.}, volume = {106}, pages = {121}, year = {1984}}

@article{CahillGlauber1969,
  title = {Density operators and quasiprobability distributions},
  author = {Cahill, K. E. and Glauber, R. J.}, journal = {Phys. Rev.},
  volume = {177}, pages = {1882}, year = {1969}}

@article{Hudson1974,
  title = {When is the {W}igner quasi-probability density non-negative?},
  author = {Hudson, R. L.}, journal = {Rep. Math. Phys.}, volume = {6},
  pages = {249}, year = {1974}}

@article{SotoClaverie1983,
  title = {When is the {W}igner function of multidimensional systems nonnegative?},
  author = {Soto, F. and Claverie, P.}, journal = {J. Math. Phys.},
  volume = {24}, pages = {97}, year = {1983}}

@article{Gross2006,
  title = {{H}udson's theorem for finite-dimensional quantum systems},
  author = {Gross, D.}, journal = {J. Math. Phys.}, volume = {47},
  pages = {122107}, year = {2006}}

@article{Mandilara2009,
  title = {Extending {H}udson's theorem to mixed quantum states},
  author = {Mandilara, A. and Karpov, E. and Cerf, N. J.},
  journal = {Phys. Rev. A}, volume = {79}, pages = {062302}, year = {2009}}

@article{Kenfack2004,
  title = {Negativity of the {W}igner function as an indicator of non-classicality},
  author = {Kenfack, A. and {\.Z}yczkowski, K.},
  journal = {J. Opt. B: Quantum Semiclassical Opt.}, volume = {6},
  pages = {396}, year = {2004}}

@article{MariEisert2012,
  title = {Positive {W}igner functions render classical simulation of quantum computation efficient},
  author = {Mari, A. and Eisert, J.}, journal = {Phys. Rev. Lett.},
  volume = {109}, pages = {230503}, year = {2012}}

@article{Chabaud2020,
  title = {Stellar representation of non-{G}aussian quantum states},
  author = {Chabaud, U. and Markham, D. and Grosshans, F.},
  journal = {Phys. Rev. Lett.}, volume = {124}, pages = {063605}, year = {2020}}

@article{Walschaers2021,
  title = {Non-{G}aussian quantum states and where to find them},
  author = {Walschaers, M.}, journal = {PRX Quantum}, volume = {2},
  pages = {030204}, year = {2021}}

@article{BoothChabaudEmeriau2022,
  title = {Contextuality and {W}igner negativity are equivalent for continuous-variable quantum measurements},
  author = {Booth, R. I. and Chabaud, U. and Emeriau, P.-E.},
  journal = {Phys. Rev. Lett.}, volume = {129}, pages = {230401}, year = {2022}}

@article{Stratonovich1956,
  title = {On distributions in representation space},
  author = {Stratonovich, R. L.}, journal = {Sov. Phys. JETP},
  volume = {31}, pages = {1012}, year = {1956}}

@article{Brif1999,
  title = {Phase-space formulation of quantum mechanics and quantum-state reconstruction for physical systems with {L}ie-group symmetries},
  author = {Brif, C. and Mann, A.}, journal = {Phys. Rev. A},
  volume = {59}, pages = {971}, year = {1999}}

@book{Perelomov1986,
  author = {Perelomov, A.},
  title = {Generalized Coherent States and their Applications},
  publisher = {Springer}, address = {Berlin}, year = {1986}}

@article{Zhang1990,
  title = {Coherent states: Theory and some applications},
  author = {Zhang, W.-M. and Feng, D. H. and Gilmore, R.},
  journal = {Rev. Mod. Phys.}, volume = {62}, pages = {867}, year = {1990}}

@book{Kam2023,
  author = {Kam, C.-F. and Zhang, W.-M. and Feng, D.-H.},
  title = {Coherent States: New Insights into Quantum Mechanics with Applications},
  series = {Lecture Notes in Physics}, volume = {1011},
  publisher = {Springer}, address = {Cham}, year = {2023},
  doi = {10.1007/978-3-031-20766-2}}

@article{Agarwal1981,
  title = {Relation between atomic coherent-state representation, state multipoles, and generalized phase-space distributions},
  author = {Agarwal, G. S.}, journal = {Phys. Rev. A}, volume = {24},
  pages = {2889}, year = {1981}}

@article{Varilly1989,
  title = {The {M}oyal representation for spin},
  author = {V{\'a}rilly, J. C. and Gracia-Bond{\'i}a, J. M.},
  journal = {Ann. Phys.}, volume = {190}, pages = {107}, year = {1989}}

@article{Klimov2017,
  title = {Generalized {SU(2)} covariant {W}igner functions and some of their applications},
  author = {Klimov, A. B. and Romero, J. L. and de Guise, H.},
  journal = {J. Phys. A: Math. Theor.}, volume = {50}, pages = {323001},
  year = {2017}}

@article{Tilma2016,
  title = {{W}igner functions for arbitrary quantum systems},
  author = {Tilma, T. and Everitt, M. J. and Samson, J. H. and Munro, W. J.
            and Nemoto, K.},
  journal = {Phys. Rev. Lett.}, volume = {117}, pages = {180401}, year = {2016}}

@article{Alonso2002,
  title = {{W}igner functions for curved spaces. {I}. On hyperboloids},
  author = {Alonso, M. A. and Pogosyan, G. S. and Wolf, K. B.},
  journal = {J. Math. Phys.}, volume = {43}, pages = {5857}, year = {2002}}

@article{Davis2021,
  title = {{W}igner negativity in spin-$j$ systems},
  author = {Davis, J. and Kumari, M. and Mann, R. B. and Ghose, S.},
  journal = {Phys. Rev. Research}, volume = {3}, pages = {033134}, year = {2021}}

@article{Davis2023,
  title = {Stellar representation of extremal {W}igner-negative spin states},
  author = {Davis, J. and Hennigar, R. A. and Mann, R. B. and Ghose, S.},
  journal = {J. Phys. A: Math. Theor.}, volume = {56}, pages = {265302},
  year = {2023}}

@article{Wodkiewicz1985,
  title = {Coherent states, squeezed fluctuations, and the {SU(2)} and {SU(1,1)} groups in quantum-optics applications},
  author = {W{\'o}dkiewicz, K. and Eberly, J. H.},
  journal = {J. Opt. Soc. Am. B}, volume = {2}, pages = {458}, year = {1985}}

@article{Gerry1985,
  title = {Dynamics of {SU(1,1)} coherent states},
  author = {Gerry, C. C.}, journal = {Phys. Rev. A}, volume = {31},
  pages = {2721}, year = {1985}}

@article{Yurke1986,
  title = {{SU(2)} and {SU(1,1)} interferometers},
  author = {Yurke, B. and McCall, S. L. and Klauder, J. R.},
  journal = {Phys. Rev. A}, volume = {33}, pages = {4033}, year = {1986}}

@article{Jing2011,
  title = {Realization of a nonlinear interferometer with parametric amplifiers},
  author = {Jing, J. and Liu, C. and Zhou, Z. and Ou, Z. Y. and Zhang, W.},
  journal = {Appl. Phys. Lett.}, volume = {99}, pages = {011110}, year = {2011}}

@article{Hudelist2014,
  title = {Quantum metrology with parametric amplifier-based photon correlation interferometers},
  author = {Hudelist, F. and Kong, J. and Liu, C. and Jing, J. and Ou, Z. Y.
            and Zhang, W.},
  journal = {Nat. Commun.}, volume = {5}, pages = {3049}, year = {2014}}

@article{Chekhova2016,
  title = {Nonlinear interferometers in quantum optics},
  author = {Chekhova, M. V. and Ou, Z. Y.},
  journal = {Adv. Opt. Photonics}, volume = {8}, pages = {104}, year = {2016}}

@article{Seyfarth2020,
  title = {{W}igner function for {SU(1,1)}},
  author = {Seyfarth, U. and Klimov, A. B. and de Guise, H. and Leuchs, G.
            and S{\'a}nchez-Soto, L. L.},
  journal = {Quantum}, volume = {4}, pages = {317}, year = {2020}}

@article{Klimov2020,
  title = {{SU(1,1)} covariant $s$-parametrized maps},
  author = {Klimov, A. B. and Seyfarth, U. and de Guise, H.
            and S{\'a}nchez-Soto, L. L.},
  journal = {J. Phys. A: Math. Theor.}, volume = {54}, pages = {065301},
  year = {2021}}

@article{Baltazar2025,
  title = {Quantum systems in the hyperbolic phase-space: Explicit maps, differential form of the star product and their applications},
  author = {Baltazar, M. and Valtierra, I. F. and Klimov, A. B.},
  journal = {Ann. Phys.}, volume = {482}, pages = {170208}, year = {2025}}

@misc{Klimov2026,
  title = {{W}igner negativity and stellar rank for {SU(1,1)} states},
  author = {Klimov, A. B. and Mu{\~n}oz, A. and Leuchs, G.
            and Gazeau, J.-P. and S{\'a}nchez-Soto, L. L.},
  note = {arXiv:2607.22810}, year = {2026}}

@book{Erdelyi1955,
  author = {Erd{\'e}lyi, A. and Magnus, W. and Oberhettinger, F.
            and Tricomi, F. G.},
  title = {Higher Transcendental Functions}, volume = {I},
  publisher = {McGraw-Hill}, address = {New York}, year = {1955}}

@misc{DLMF,
  title = {{NIST} Digital Library of Mathematical Functions, Chap.~14},
  note = {\url{http://dlmf.nist.gov/}}}

\end{document}